\documentclass{ws-procs9x6}
\begin{document}

\title{LIGHT-FRONT DENSITIES\\
FOR TRANSVERSELY POLARIZED HADRONS}

\author{C. LORC\'E$^*$ and M. VANDERHAEGHEN}

\address{Institut fuer Kernphysik, Johannes Gutenberg-Universitaet,\\
Mainz, D55099, Germany\\
$^*$E-mail: lorce@kph.uni-mainz.de}

\author{B. PASQUINI}

\address{Dipartimento di Fisica Nucleare e Teorica, Universit\`a degli Studi di Pavia, and\\
INFN, Sezione di Pavia, I-27100 Pavia, Italy\\
E-mail: Barbara.Pasquini@pv.infn.it}

\begin{abstract}
We discuss the recent interpretation of quark-distribution functions in the 
plane transverse to the light-cone direction. Such a mapping is model independent and allows one to build up multidimensional pictures of the hadron and to develop a semi-classical interpretation of the quark dynamics. We comment briefly the results obtained from the form factors of the nucleon. A generalization to a target with arbitrary spin led to a set of preferred values for the electromagnetic coupling characterizing structureless particles. Generalized polarizabilities can also be interpreted in that frame as the distortion of the charge densities due to a quasi-static external electromagnetic field. Finally, we present 
results for the generalized transverse-momentum dependent distributions which encode the most complete information about quark distributions.
\end{abstract}


\bodymatter

\section{Introduction}
Hadrons are composite objects. Their interaction with external probes like \emph{e.g.} photons is parameterized in terms of Lorentz-invariant functions. The latter encode information about the distribution of partons in momentum and/or position space. In order to extract this information, one has to go to a frame where the hadron moves with almost the speed of light allowing for a (quasi-)probabilistic/charge-density interpretation. This interpretation is in principle model independent. A brief discussion can be found in Section~\ref{Sec2}.

We first focus on form factors (FFs) which, by Fourier transform, give information on the charge distribution in the transverse plane, or impact parameter space, see Section~\ref{Sec3}. While for a longitudinally polarized target the pattern is axially symmetric, two-dimensional multipoles appear for a transversely polarized target. Such multipoles turn out to be a manifestation of the target composite nature and to be intimately connected to quark orbital angular momentum. Since a structureless particle does not contain any intrinsic orbital angular momentum, it should not show any multipole pattern. This constraint then leads to a set of preferred values for the electromagnetic (EM) coupling (``natural'' 
EM moments) of particles with arbitrary spin.

Interpretation of generalized polarizabilities (GPs) is then discussed in Section~\ref{Sec4} in the infinite momentum frame. 
They describe how the charge distribution is affected by a quasi-static external electromagnetic field. Similarly to FFs, interesting multipole patterns also show up. 
However, working in the infinite momentum frame we can arrive to a spatial representation for only three combinations of the six independent GPs.

Finally, we present in Section~\ref{Sec5} the recently introduced generalized transverse-momentum dependent distributions (GTMDs) and their connection with the  generalized parton distributions (GPDs), transverse-momentum dependent distributions (TMDs) and Wigner distributions. The last ones encode all the possible correlations between quark transverse position and  momentum. As it is not possible to directly access Wigner distributions from experiments, we use relativistic quark models to learn about the quark dynamics encoded in these quantities.

\section{Impact parameter space}\label{Sec2}
According to the standard interpretation [\refcite{Ernst:1960zza,Sachs:1962zzc}] the charge density can be identified with the three-dimensional Fourier transform of the electric Sachs form factor $G_E$, i.e.
\begin{equation}
\rho(\vec r)=\int\frac{\textrm{d}^3q}{(2\pi)^3}\,e^{-i\vec q\cdot\vec r}\,G_E(\vec q),
\label{eq:rho}
\end{equation}
where $G_E$ is defined as the matrix element of the charge current between the initial and final nucleon with momentum $\vec P$ and $\vec{P}'=\vec{P}+\vec{q}$, respectively.
The identification in Eq.~\eqref{eq:rho} is valid only in the nonrelativistic approximation. To work out the Fourier transform, one has to know the FFs for every $Q^2$. In the Breit frame the latter is identified with the three-momentum of the virtual photon, i.e. $Q^2=\vec q\,^2$. This means that for every value of $Q^2$ we have to move to a different frame and the charge density undergoes naturally a different Lorentz contraction. Moreover, in order to have a probabilistic/charge-density interpretation, the number of particles should be conserved. However, in the Breit frame nothing prevents the photon to create or annihilate a quark-antiquark pair.

All these problems are cured in the infinite momentum frame (IMF) with $q^+=q^0+q^z=0$ (the so-called Drell-Yan-West frame). In such a frame the photon is kinematically not allowed to change the number of quarks since the light-cone momentum of a massive particle is strictly positive $p^+>0$. Moreover, in the limit $p^+\rightarrow\infty$ the hadron undergoes an extreme Lorentz contraction. Only a two-dimensional charge density [\refcite{Soper:1976jc,Burkardt:2000za,Burkardt:2002hr}] is then meaningful and can be identified with the two-dimensional Fourier transform of the matrix element of $J^+$
\begin{equation}
\rho_{\vec s}(\vec b)=\int\frac{\textrm{d}^2q_\perp}{(2\pi)^2}\,\frac{e^{-i\vec q_\perp\cdot\vec b}}{2p^+}\,\langle p^+,\tfrac{\vec q_\perp}{2},\vec s|J^+(0)|p^+,-\tfrac{\vec q_\perp}{2},\vec s\rangle,
\end{equation}
with the photon virtuality $Q^2=\vec q_\perp\,\!\!\!\!^2$, and $\vec s$ denoting the hadron polarization.

\section{Form factors}\label{Sec3}

For a spin-$1/2$ hadron with definite light-cone helicity $\lambda$, the charge density is simply given by the Fourier transform of the Dirac FF
\begin{equation}
\rho_\lambda(\vec b)=\int_0^\infty\frac{\textrm{d}Q}{2\pi}\,Q\,J_0(bQ)\,F_1(Q^2),
\end{equation}
where $J_0$ is the cylindrical Bessel function. Since no preferred direction appears in the transverse plane, it comes without any surprise that the corresponding charge density is axially symmetric. Using phenomenological parameterizations of the experimental nucleon FFs, one observes a counterintuitive negative core in the neutron charge distribution [\refcite{Miller:2007uy}]. For a transverse polarization, the charge density receives a dipole contribution [\refcite{Carlson:2007xd}] from the Fourier transform of the Pauli FF
\begin{equation}
\rho_{s_\perp}(\vec b)=\rho_\lambda(\vec b)+\sin\phi_b\int_0^\infty\frac{\textrm{d}Q}{2\pi}\,\frac{Q^2}{2M}\,J_1(bQ)\,F_2(Q^2),
\label{eq:pauli}
\end{equation}
where $\phi_b$ is the angle between quark transverse position $\vec b$ and transverse hadron polarization $\vec s_\perp$. This transverse polarization gives a preferred direction in the transverse plane allowing for multipole patterns. It also involves matrix elements with helicity flip, where quark orbital angular momentum is necessary. The dipole moment in natural units ($e/2M$) is given by $F_2(0)$, \emph{i.e.} the anomalous magnetic moment. Higher multipole moments can be observed for hadrons with higher spin [\refcite{Carlson:2008zc,Alexandrou:2008bn,Alexandrou:2009hs,Lorce:2009bs}]. There is therefore a connection between anomalous EM moments, quark orbital angular momentum and distortions of the charge densities. 
In the case of a particle \emph{without} internal structure, there can not be any intrinsic orbital angular momentum and the charge densities remain axially symmetric. This provides us with a sufficient condition to derive the natural values of the EM moments [\refcite{Lorce:2009bs}] characterizing the interaction of a structureless particle of any spin with the electromagnetic field. The general results agree with the known lower spin cases (Standard Model and Supergravity) and support the prediction of a $g=2$ gyromagnetic ratio for any spin.
\begin{table}[h]
\tbl{Natural EM moments of a spin-$j$ particle with electric charge $Z=+1$ are organized according to a pseudo Pascal triangle, when expressed in terms of natural units of $e/M^l$ and $e/2M^l$ for $G_{El}(0)$ and $G_{Ml}(0)$, respectively.}
{\begin{tabular}{@{}cccccccc@{}}\toprule
$j$&$G_{E0}(0)$&$G_{M1}(0)$&$G_{E2}(0)$&$G_{M3}(0)$&$G_{E4}(0)$&$G_{M5}(0)$&$G_{E6}(0)$\\
\colrule
$0$&$1$&$0$&$0$&$0$&$0$&$0$&$0$\\
$1/2$&$1$&$1$&$0$&$0$&$0$&$0$&$0$\\
$1$&$1$&$2$&$-1$&$0$&$0$&$0$&$0$\\
$3/2$&$1$&$3$&$-3$&$-1$&$0$&$0$&$0$\\
$2$&$1$&$4$&$-6$&$-4$&$1$&$0$&$0$\\
$5/2$&$1$&$5$&$-10$&$-10$&$5$&$1$&$0$\\
$3$&$1$&$6$&$-15$&$-20$&$15$&$6$&$-1$\\
\botrule
\end{tabular}}
\label{EMmoments}
\end{table}

\section{Generalized Polarizabilities}\label{Sec4}

In the presence of an external electric field, any charge density is deformed and develops an induced polarization.
 The ``ease'' by which such a charge density is deformed is described by the electric polarizability. Virtual Compton scattering (VCS) allows us to study [\refcite{Gorchtein:2009qq}] the distortion of transverse charge densities in an external electric field. The nucleon VCS tensor is given by
\begin{equation}
H^{\mu\nu}=-i\int\textrm{d}^4x\,e^{-iq\cdot x}\,\langle p',\vec s_N|T\{J^\mu(x)J^\nu(0)\}|p,\vec s_N\rangle.
\end{equation}
In the limit of low energy of the final real photon, it is described in terms of six Generalized Polarizabilities (GPs). To obtain a spatial representation, we consider the VCS process in a light-front frame. In the low-energy limit $\nu=q\cdot P/M\to 0$ ($P=\tfrac{p'+p}{2}$) the VCS amplitude $H=\epsilon'^*_\nu\,H^{+\nu}$ describes the linear response to an external quasi-static electric dipole field, allowing us to define an induced polarization $\vec P(\vec q)$
\begin{equation}
i\vec\epsilon'^*_\perp\cdot\vec P(\vec q)\equiv\frac{1+\tau}{2P^+}\,\frac{\partial H}{\partial\nu}\Big|_{\nu=0}.
\end{equation}
By Fourier transforming $\vec P(\vec q)$ we obtain the spatial distortion of the charge density induced by the external electric field ($\hat a=\vec a/|\vec a|$), i.e.
\begin{equation}
\begin{split}
\vec P(\vec b)&=\int\frac{\textrm{d}^2q_\perp}{(2\pi)^2}\,e^{-i\vec q_\perp\cdot\vec b}\,\vec P(\vec q)\\
&=\hat b\int_0^\infty\frac{\textrm{d}Q}{2\pi}\,Q\,J_1(bQ)\,A(Q^2)\\
&\phantom{=}-\hat b\left(\vec s_\perp\times\hat e_z\right)\cdot\hat b\int_0^\infty\frac{\textrm{d}Q}{2\pi}\,Q\,J_2(bQ)\,B(Q^2)\\
&\phantom{=}+\left(\vec s_\perp\times\hat e_z\right)\int_0^\infty\frac{\textrm{d}Q}{2\pi}\,Q\left[J_0(bQ)\,C(Q^2)+\frac{J_1(bQ)}{bQ}\,B(Q^2)\right].
\end{split}
\end{equation}
Only three multipoles are at our disposal to interpret six GPs. The first term corresponds just to the standard dipole distortion of the charge density, irrespective of the nucleon polarization. The other two terms appear only when the nucleon is transversely polarized. In absence of external electric field and for a transversely polarized nucleon, the charge distribution already contains a dipole structure (see Eq.~\eqref{eq:pauli}). Applying an external electric field on this dipole charge density then leads to a monopole and a quadrupole distortions.

\section{Generalized Transverse-Momentum dependent Distributions}\label{Sec5}

FFs provide information about the quark transverse position while parton distribution functions (PDFs) give their longitudinal-momentum distribution. The full correlation between quark transverse position and longitudinal momentum is contained in GPDs leading then to a three-dimensional picture of the nucleon. A dual three-dimensional picture of the nucleon is provided by TMDs which give the full quark three-momentum distribution. The full correlation between quark three-momentum and transverse position is captured by the Wigner distributions defined as Fourier transform in the transverse plane 
of the so-called GTMDs. All FFs, PDFs, GPDs, and TMDs appear just as particular limits and/or projections of GTMDs. There are sixteen GTMDs [\refcite{Meissner:2009ww}] which parameterize the general parton correlator
\begin{equation}
W^{[\Gamma]}_{\lambda'\lambda}=\frac{1}{2}\int\frac{\textrm{d}^4z}{(2\pi)^3}\,\delta(z^+)\,e^{iq\cdot z}\langle p',\lambda'|\overline\psi(-\tfrac{1}{2}z)\Gamma\mathcal W(-\tfrac{1}{2}z,\tfrac{1}{2}z|n)\psi(\tfrac{1}{2}z)|p,\lambda\rangle,
\end{equation}
where $\Gamma$ is a matrix in Dirac space, $\psi$ is the quark field, and $\mathcal W$ a Wilson line which preserves gauge invariance. These GTMDs $X(x,\xi,\vec k_\perp^2,\vec k_\perp\cdot\vec\Delta_\perp,\vec\Delta_\perp^2;\eta)$ are functions of quark mean momentum $(x,\vec k_\perp)$ and momentum transfer $(\xi,\vec\Delta_\perp)$. The parameter $\eta$ indicates whether the Wilson line is past- or future-pointing.

Wigner Distributions in the context of quantum field theory have already been discussed to some extent in the Breit frame [\refcite{Ji:2003ak,Belitsky:2003nz}]. For the reason mentioned before, it is preferable to work in the IMF. The  distribution of an unpolarized quark in an unpolarized hadron is given by the GTMD $F^e_{11}$, and is related to the GPD $H$ and the TMD $f_1$ as follows
\begin{equation}
\begin{split}
H(x,0,\vec\Delta_\perp^2)&=\int\textrm{d}^2k_\perp\,F^e_{11}(x,0,\vec k_\perp^2,\vec k_\perp\cdot\vec\Delta_\perp,\vec\Delta_\perp^2),\\
f_1(x,\vec k_\perp^2)&=F^e_{11}(x,0,\vec k_\perp^2,0,0).
\end{split}
\end{equation}
By Fourier transforming $F^e_{11}$ with respect to $\vec\Delta_\perp$ and integrating over $x$, we obtain a (transverse) phase-space distribution $\rho(\vec k_\perp^2,\vec k_\perp\cdot\vec b,\vec b^2)$. The only two available transverse vectors are $\vec k_\perp$ and $\vec b$. This means that for fixed $\hat k_\perp\cdot\hat b$ and $|\vec k_\perp|$ the distribution is axially symmetric, which 
is physically meaningful since a global rotation of the distribution around 
$\hat e_z$ should not have any effect. 

\begin{figure}
\begin{center}
\begin{tabular}{c@{\hspace{1cm}}c}
\psfig{file=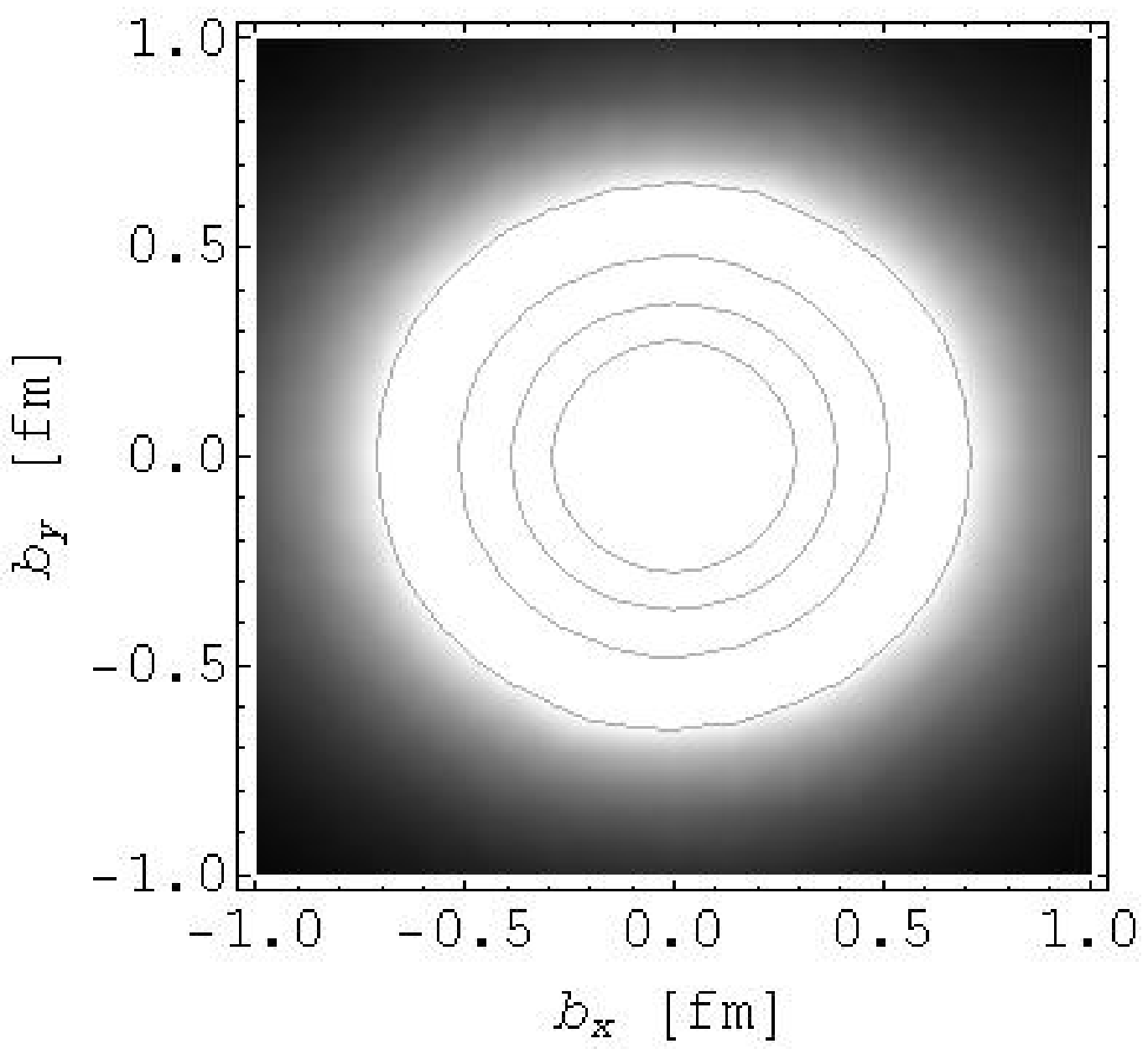,width=5cm}&\psfig{file=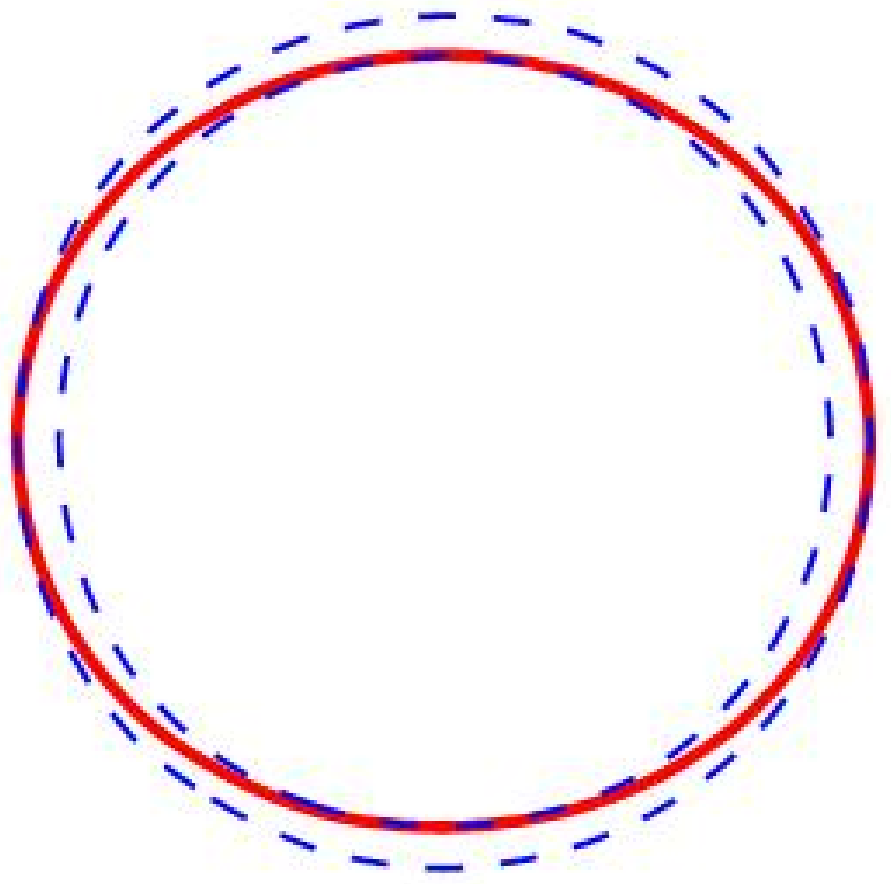,width=4cm}
\end{tabular}
\end{center}
\caption{Distribution (left) and equi-amplitude curve (right) in the transverse plane with $\hat k_\perp=\hat e_y$ and $|\vec k_\perp|=0.45$ GeV  for an unpolarized quark in an unpolarized nucleon.}
\label{aba:fig1}
\end{figure}

If we are now interested in the amplitude to find a fixed transverse momentum $\vec k_\perp$ in the transverse plane, the distribution is not axially symmetric anymore due to the $\vec k_\perp\cdot\vec b$ dependence. This amplitude can not be directly accessed from experiments. We therefore use successful phenomenological constituent quark models (namely the Light-Cone Constituent Quark Model [\refcite{Pasquini:2007xz}] and the Chiral Quark-Soliton Model [\refcite{Lorce:2007fa}]) restricted to the valence sector [\refcite{LorcePasquini}] to compute $F^e_{11}$. Both models give the same qualitative picture with a larger distribution amplitude when $\vec k_\perp\perp\vec b$ and smaller when $\vec k_\perp\parallel\vec b$, see Fig. \ref{aba:fig1}. 
This can be understood with naive semi-classical arguments. The radial momentum $(\vec k_\perp\cdot\hat b)\,\hat b$ of a quark has to decrease rapidly in the periphery because of confinement. The polar momentum $\vec k_\perp-(\vec k_\perp\cdot\hat b)\,\hat b$ receives a contribution from the orbital motion of the quark which can still be significant in the periphery (in an orbital motion, one does not need to reduce the momentum to avoid a quark escape). This naive picture also tells us that this phenomenon should become more pronounced as we go to peripheral regions ($|\vec b|\gg$) and to high quark momenta ($|\vec k_\perp|\gg$). This tendency is supported by both models.

\begin{figure}
\begin{center}
\psfig{file=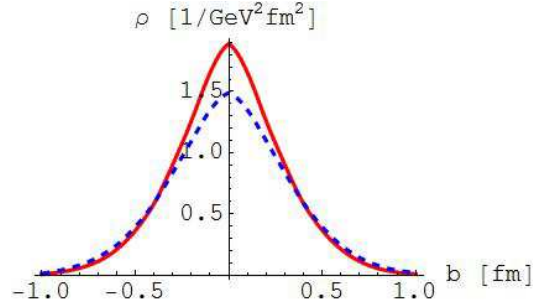,width=7cm}
\end{center}
\caption{Radial distributions in the transverse plane with $\hat k_\perp\cdot\hat b=1$ and $|\vec k_\perp|=0.55$ GeV for an unpolarized quark in an unpolarized proton. Solid curve corresponds to half of the up quark distribution. Dashed curve corresponds to the down quark distribution.}
\label{aba:fig2}
\end{figure}

It is also interesting to compare up and down quark distributions. Let us fix once again $\hat k_\perp\cdot\hat b$ and $|\vec k_\perp|$. The distribution being axially symmetric in the transverse plane, we focus on the radial distribution, see Fig. \ref{aba:fig2}.
The up quark distribution has been divided by two for comparison with the down quark distribution. Up quarks appear to be more concentrated around the center than down quarks. For a neutron, we just have to exchange up and down quarks. We can therefore see that the center of the neutron is negative, in agreement with the conclusion obtained using the phenomenological neutron FFs.

\section{Summary}

The standard interpretation of FFs in the Breit frame is not fully consistent with the principles of quantum mechanics and relativity. However, a probabilistic/charge-density interpretation is available in a frame where the target is moving with almost the speed of light. It then turns out that FFs describe the distribution of quark in the transverse plane. While this distribution is axially symmetric for a longitudinally polarized target, multipoles appear when the target is transversely polarized. This distortion comes from the intrinsic orbital angular momentum and hence the composite nature of the target. This observation allowed us to derive the natural EM moments characterizing the interaction of any structureless particle with the electromagnetic field. We also tried to interpret the Generalized Polarizabilities in a light-front frame based on the VCS process in the low-energy limit. It appeared that only three combinations out of six GPs can be mapped out. Finally, we discussed briefly phase-space (Wigner) distribution using constituent quark models. Non-trivial pattern have been obtained and interpreted semi-classically as related to orbital angular momentum. 

\section*{Acknowledgments}

This work was supported in part  by
the Research Infrastructure Integrating Activity
``Study of Strongly Interacting Matter'' (acronym HadronPhysics2, Grant
Agreement n. 227431) under the Seventh Framework Programme of the
European Community, by the Italian MIUR through the PRIN 
2008EKLACK ``Structure of the nucleon: transverse momentum, transverse 
spin and orbital angular momentum''.

\bibliographystyle{ws-procs9x6}
\bibliography{Lorce-v2}

\begin{thebibliography}{10}

\bibitem{Ernst:1960zza}
F.~J. Ernst, R.~G. Sachs and K.~C. Wali, {\em Phys. Rev.} {\bf 119}, 1105
  (1960).

\bibitem{Sachs:1962zzc}
R.~G. Sachs, {\em Phys. Rev.} {\bf 126}, 2256 (1962).

\bibitem{Soper:1976jc}
D.~E. Soper, {\em Phys. Rev. D} {\bf 15}, 1141 (1977).

\bibitem{Burkardt:2000za}
M.~Burkardt, {\em Phys. Rev. D} {\bf 62}, 071503 (2000).

\bibitem{Burkardt:2002hr}
M.~Burkardt, {\em Int. J. Mod. Phys. A} {\bf 18}, 173 (2003).

\bibitem{Miller:2007uy}
G.~A. Miller, {\em Phys. Rev. Lett.} {\bf 99}, 112001 (2007).

\bibitem{Carlson:2007xd}
C.~E. Carlson and M.~Vanderhaeghen, {\em Phys. Rev. Lett.} {\bf 100}, 032004
  (2008).

\bibitem{Carlson:2008zc}
C.~E. Carlson and M.~Vanderhaeghen, {\em Eur. Phys. J. A} {\bf 41}, 1 (2009).

\bibitem{Alexandrou:2008bn}
C.~Alexandrou{\it\mbox{ }et al.}, {\em Phys. Rev. D} {\bf 79}, 014507 (2009).

\bibitem{Alexandrou:2009hs}
C.~Alexandrou{\it\mbox{ }et al.}, {\em Nucl. Phys. A} {\bf 825}, 115 (2009).

\bibitem{Lorce:2009bs}
C.~Lorce, {\em Phys. Rev. D} {\bf 79}, 113011 (2009).

\bibitem{Gorchtein:2009qq}
M.~Gorchtein, C.~Lorce, B.~Pasquini and M.~Vanderhaeghen, {\em Phys. Rev.
  Lett.} {\bf 104}, 112001 (2010).

\bibitem{Meissner:2009ww}
S.~Meissner, A.~Metz and M.~Schlegel, {\em JHEP} {\bf 0809}, p. 056 (2009).

\bibitem{Ji:2003ak}
X.~D. Ji, {\em Phys. Rev. Lett.} {\bf 91}, 062001 (2003).

\bibitem{Belitsky:2003nz}
A.~V. Belitsky, X.~D. Ji and F.~Yuan, {\em Phys. Rev. D} {\bf 69}, 074014
  (2004).

\bibitem{Pasquini:2007xz}
B.~Pasquini and S.~Boffi, {\em Phys. Lett. B} {\bf 653}, 23 (2007).

\bibitem{Lorce:2007fa}
C.~Lorce, {\em Phys. Rev. D} {\bf 79}, 074027 (2009).

\bibitem{LorcePasquini}
C.~Lorce and B.~Pasquini, Wigner distributions on the light cone, in
  preparation, (2010).

\end{thebibliography}

\end{document}